\documentclass[aip,jcp,reprint,noshowkeys]{revtex4-1}
\usepackage{graphicx,dcolumn,bm,xcolor,microtype,multirow,amscd,amsmath,amssymb,amsfonts,physics,mhchem}

\usepackage{tikz}
\usetikzlibrary{arrows,positioning,shapes.geometric}
\usetikzlibrary{decorations.pathmorphing}

\tikzset{snake it/.style={
decoration={snake, 
    amplitude = .4mm,
    segment length = 2mm},decorate}
}

\newcommand{\hH}{\Hat{H}}

\newcommand{\mD}{\mathcal{D}}
\newcommand{\mA}{\mathcal{A}}

\newcommand{\ka}{\ket{\alpha}}
\newcommand{\kI}{\ket{I}}

\usepackage{mathpazo,libertine}
\usepackage[normalem]{ulem}
\definecolor{darkgreen}{RGB}{0, 180, 0}

\newcommand{\mc}{\multicolumn}
\newcommand{\tabc}[1]{\multicolumn{1}{c}{#1}}
\newcommand{\fnm}{\footnotemark}
\newcommand{\fnt}{\footnotetext}

\newcommand{\SI}{\textcolor{blue}{supplementary material}}
\newcommand{\tX}{\text{X}}

\newcommand{\EsCI}{E_\text{sCI}}
\newcommand{\EDMC}{E_\text{DMC}}
\newcommand{\EexFCI}{E_\text{exFCI}}
\newcommand{\EexDMC}{E_\text{exDMC}}

\newcommand{\ex}[6]{$^{#1}#2_{#3}^{#4}(#5 \rightarrow #6)$}
\newcommand{\exc}[4]{$^{#1}#2_{#3}^{#4}$}

\newcommand{\IneV}[1]{#1 eV}

\newcommand{\PsiT}{\Psi_\text{T}}

\newcommand{\Nst}{N_\text{states}}

\newcommand{\Ndet}{N_\text{det}}

\newcommand{\LCPQ}{Laboratoire de Chimie et Physique Quantiques, Universit\'e de Toulouse, CNRS, UPS, France}
\newcommand{\CSD}{Computational Science Division, Argonne National Laboratory, Argonne, IL 60439, United States
of America}
\newcommand{\CEISAM}{Laboratoire CEISAM - UMR CNRS 6230, Universit\'e de Nantes, 2 Rue de la Houssini\`ere, BP 92208, 44322 Nantes Cedex 3, France}

\begin{document}	

\title{\textcolor{blue}{\textbf{Influence of Pseudopotentials on Excitation Energies From Selected Configuration Interaction and Diffusion Monte Carlo}}}

\author{Anthony Scemama}
\affiliation{\LCPQ}
\author{Michel Caffarel}
\affiliation{\LCPQ}
\author{Anouar Benali}
\affiliation{\CSD}
\author{Denis Jacquemin}
\affiliation{\CEISAM}    
\author{Pierre-Fran{\c c}ois Loos}
\email[Corresponding author: ]{loos@irsamc.ups-tlse.fr}
\affiliation{\LCPQ}

\begin{abstract}
Due to their diverse nature, the faithful description of excited states within electronic structure theory methods remains one of the grand challenges of modern theoretical chemistry.
Quantum Monte Carlo (QMC) methods have been applied very successfully to ground state properties but still remain generally less effective than other non-stochastic methods for electronically excited states.
Nonetheless, we have recently reported accurate excitation energies for small organic molecules at the fixed-node diffusion Monte Carlo (FN-DMC) within a Jastrow-free QMC protocol relying on a deterministic and systematic construction of nodal surfaces using the selected configuration interaction (sCI) algorithm known as CIPSI (Configuration Interaction using a Perturbative Selection made Iteratively).
Albeit highly accurate, these all-electron calculations are computationally expensive due to the presence of core electrons. 
One very popular approach to remove these chemically-inert electrons from the QMC simulation is to introduce pseudopotentials (also known as effective core potentials).
Taking the water molecule as an example, we investigate the influence of Burkatzki-Filippi-Dolg (BFD) pseudopotentials and their associated basis sets on vertical excitation energies obtained with sCI and FN-DMC methods.
Although these pseudopotentials are known to be relatively safe for ground state properties, we evidence that special care may be required if one strives for highly accurate vertical transition energies.
Indeed, comparing all-electron and valence-only calculations, we show that using pseudopotentials with the associated basis sets can induce differences of the order of 0.05 eV on the excitation energies. 
Fortunately, a reasonable estimate of this shift can be estimated at the sCI level.
\end{abstract}


\maketitle

\section{Introduction}
At the very heart of photochemistry lies the subtle role played by low-lying electronic states and their mutual interactions.\cite{Delgado_2010,Palczewski_2006,Bernardi_1996,Olivucci_2010,Robb_2007}
In general, the correct description of these phenomena requires to locate with enough accuracy the first few low-lying excited states of the system and to understand how such states interact not only between themselves (conical intersections, spin-orbit effects, \ldots) but also with other degrees of freedom (coupling with ro-vibrational modes, environment effects, \ldots).
For example, in the case of the photophysics of vision, precious information can be gained by exploring the excited states of polyenes \cite{Serrano-Andres_1993,Cave_1988b,Lappe_2000,Maitra_2004,Cave_2004,Wanko_2005,Starcke_2006,Angeli_2010,Mazur_2011,Huix-Rotllant_2011} that are closely related to rhodopsin
which is involved in visual phototransduction. \cite{Gozem_2014,Huix-Rotllant_2010,Xu_2013,Schapiro_2014,Tuna_2015,Manathunga_2016}

Accurate and efficient electronic structure methods are now available for the computation of molecular excited states.
Time-dependent density-functional theory (TD-DFT) \cite{Casida} is undoubtedly at the front of the pack thanks to its favorable cost/accuracy ratio, although several well-documented shortcomings have been put forward in the past twenty years. \cite{Woodcock_2002,Tozer_2003,Tozer_1999,Dreuw_2003,Sobolewski_2003,Dreuw_2004,Maitra_2017,Tozer_1998,Tozer_2000,Casida_1998,Casida_2000,Tapavicza_2008,Levine_2006,Elliott_2011}.
More expensive methods, such as CIS(D), \cite{Head-Gordon_1995} CC2, \cite{Hattig_2000} CC3, \cite{Koch_1997} ADC(2), \cite{Dreuw_2015} ADC(3), \cite{Harbach_2014} EOM-CCSD \cite{Purvis_1982} (and higher orders CC approaches \cite{Kucharski_1991}) are also available. 
Albeit often more computationally expensive, one can also rely on multiconfigurational methods such as the complete active space self-consistent field (CASSCF) method, \cite{Roos} its second-order perturbation-corrected variant (CASPT2), \cite{Andersson_1990} as well as the second-order $n$-electron valence state perturbation theory (NEVPT2), \cite{Angeli_2001a} to compute accurate transition energies.
Alternatively to the mainstream methods mentioned above, selected configuration interaction (sCI) methods \cite{Bender_1969,Whitten_1969,Huron_1973,Evangelisti_1983} have demonstrated to be valuable alternatives for the computation of highly accurate transition energies for small molecules. \cite{Giner_2013,Caffarel_2014,Giner_2015,Garniron_2017b,Caffarel_2016,Holmes_2016,Sharma_2017,Holmes_2017,Scemama_2018a,Scemama_2018b,Loos_2018b,Garniron_2018,Evangelista_2014,Schriber_2016,Zimmerman_2017,Loos_2019,Garniron_2019}

Pushing further this idea, we have reported, in a recent study, \cite{Scemama_2018b} accurate excitation energies for two small organic molecules (water and formaldehyde) using fixed-node diffusion Monte Carlo (FN-DMC) \cite{Kalos_1974, Ceperley_1979, Reynolds_1982, Foulkes_1999, Lester_2009, Austin_2012} within a Jastrow-free quantum Monte Carlo (QMC) protocol relying on a deterministic and systematic construction of nodal surfaces using the sCI algorithm known as CIPSI (Configuration Interaction using a Perturbative Selection made Iteratively). \cite{Huron_1973,Giner_2013,Giner_2015, Caffarel_2014,Scemama_2014,Caffarel_2016,Scemama_2016,Garniron_2017b,Scemama_2018a,Scemama_2018b,Dash_2018,Garniron_2019}. 
Within FN-DMC, ensuring accurate calculations of vertical transition energies is far from being straightforward \cite{Grossman_2001, Porter_2001, Porter_2001a, Puzder_2002, Williamson_2002, Aspuru-Guzik_2004, Schautz_2004, Bande_2006, Bouabca_2009, Purwanto_2009, Zimmerman_2009, Dubecky_2010, Send_2011, Guareschi_2013, Guareschi_2014, Dupuy_2015, Zulfikri_2016, Guareschi_2016a, Blunt_2017, Robinson_2017, Shea_2017, Zhao_2016, Scemama_2018a, Scemama_2018b} as the mechanism and degree of error compensation of the fixed-node error \cite{Ceperley_1991, Bressanini_2001, Rasch_2012, Rasch_2014, Kulahlioglu_2014} in the ground and excited states are mostly unknown, expect in a few cases. \cite{Bajdich_2005, Bressanini_2008, Scott_2007, Bressanini_2005a, Bressanini_2005b, Bressanini_2012, Mitas_2006, Scott_2007, Loos_2015}
However, our study has clearly evidenced that the fixed-node errors in the ground and excited states obtained with sCI trial wave functions cancel out to a large extent, allowing for the determination of accurate vertical excitation energies for both the singlet and triplet manifolds.

The FN-DMC results reported in Ref.~\onlinecite{Scemama_2018b} are based on all-electron calculations, i.e., we do not use pseudopotentials (also known as pseudopotentials) to model the core electrons, contrary to what is done in most QMC calculations on large systems. \cite{Austin_2012, Dubecky_2016, Benali_2014, Ambrosetti_2014}
Our motivation was to avoid any unnecessary approximation on our excitation energies.
However, due to the large fluctuations associated with the very energetic core electrons, all-electron calculations are computationally expensive and must be avoided for large systems.
It is then highly desirable to quantify the error that one introduces with pseudopotentials.
This problem is investigated here both for sCI and DMC calculations using the water molecule as a test system.

This manuscript is organized as follows. 
The CIPSI algorithm used to obtain ground and excited-state wave functions is presented in Sec.~\ref{sec:CIPSI}.
Computational details are reported in Sec.~\ref{sec:comp_details}. 
In Sec.~\ref{sec:res}, we discuss our results and we draw our conclusions in Sec.~\ref{sec:ccl}. 
Unless otherwise stated, atomic units are used throughout this study.

\begin{figure}
	\includegraphics[width=\linewidth]{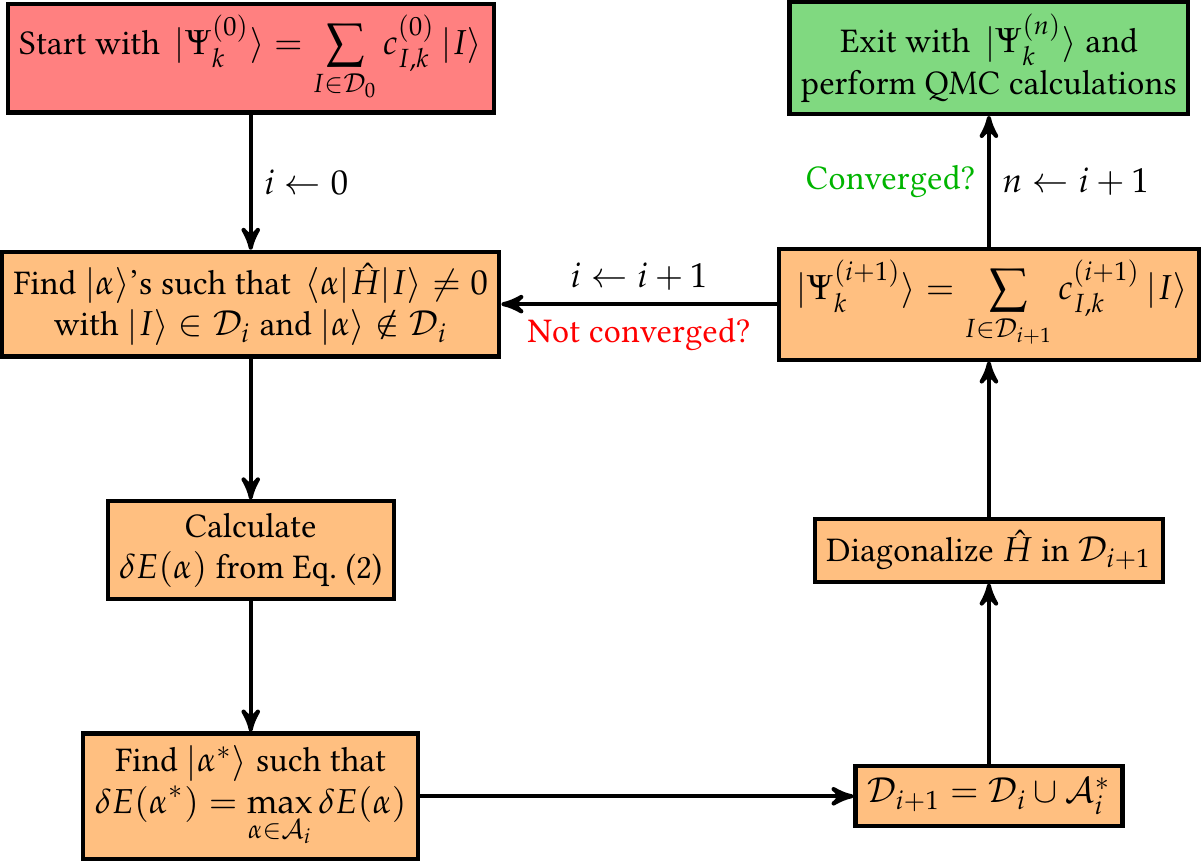}
\caption{
\label{fig:algo}
The CIPSI algorithm. See text for notations.}
\end{figure}

\begin{squeezetable}
\begin{table*}
\caption{
\label{tab:H2O}
Vertical excitation energies (in eV) for the three lowest singlet and three lowest triplet excited states of water obtained with all-electron AVXZ basis set and with the combination of BFD pseudopotentials and valence-only AVXZ basis sets (X $=$ D, T, and Q).
The error bar corresponding to one standard error is reported in parenthesis.
The relative difference between the all-electron and the corresponding pseudopotential calculation is reported in square brackets.}
\begin{ruledtabular}
\begin{tabular}{llllllll}
	Basis		&	Method				&	\mc{3}{c}{Singlet excitations}	&	\mc{3}{c}{Triplet excitations}	\\
											\cline{3-5}							\cline{6-8}
				&						&	\tabc{\ex{1}{B}{1}{}{n}{3s}} 	&	\tabc{\ex{1}{A}{2}{}{n}{3p}} 	&	\tabc{\ex{1}{A}{1}{}{n}{3s}} 	
										&	\tabc{\ex{3}{B}{1}{}{n}{3s}} 	&	\tabc{\ex{3}{A}{2}{}{n}{3p}} 	&	\tabc{\ex{3}{A}{1}{}{n}{3s}} 	\\
	
\hline
	AVDZ		&	exFCI\fnm[1]		&	7.53			&	9.32			&	9.94			&	7.14			&	9.14			&	9.48			\\
				&	SHCI\fnm[2]			&					&					&	9.94(1)			&					&					&					\\
				&	exDMC\fnm[1]		&	7.73(1)			&	9.48(1)			&	10.10(1)		&	7.36(1) 		&	9.33(1)			&	9.63(1)			\\
	AVDZ-BFD	&	exFCI\fnm[3]		&	7.48[-0.05]		&	9.28[-0.04]		&	9.88[-0.06]		&	7.07[-0.07]		&	9.11[-0.03]		&	9.43[-0.05]		\\
				&	SHCI\fnm[2]			&					&					&	9.86(1)[-0.08]	&					&					&					\\
				&	exDMC\fnm[3]		&	7.65(1)[-0.08]	&	9.45(1)[-0.03]	&	10.00(1)[-0.10]	&	7.26(1)[-0.10]	&	9.27(1)[-0.06]	&	9.54(1)[-0.09]	\\
				&	DMC\{J,O\}\fnm[1]	&					&					&	9.97(1)			&					&					&					\\
\hline
	AVTZ		&	exFCI\fnm[1]		&	7.63			&	9.41			&	9.99			&	7.25			&	9.24			&	9.54			\\
				&	SHCI\fnm[2]			&					&					&	10.00(0)		&					&					&					\\
				&	exDMC\fnm[1]		&	7.70(2)			&	9.47(2)			&	10.05(2)		&	7.35(1)			&	9.32(1)			&	9.61(1)			\\
	AVTZ-BFD	&	exFCI\fnm[3]		&	7.58[-0.05]		&	9.38[-0.03]		&	9.93[-0.06]		&	7.16[-0.09]		&	9.21[-0.03]		&	9.47[-0.07]		\\
				&	SHCI\fnm[2]			&					&					&	9.93(1)[-0.07]	&					&					&					\\
				&	exDMC\fnm[3]		&	7.66(1)[-0.04]	&	9.49(1)[+0.02]	&	10.04(1)[-0.01]	&	7.25(1)[-0.10]	&	9.30(1)[-0.02]	&	9.55(1)[-0.06]	\\
				&	DMC\{J,O\}\fnm[2]	&					&					&	10.01(1)		&					&					&					\\
\hline
	AVQZ		&	exFCI\fnm[1]		&	7.68			&	9.46			&	10.03			&	7.30			&	9.29			&	9.58			\\
				&	SHCI\fnm[2]			&					&					&	10.02(1)		&					&					&					\\
				&	exDMC\fnm[1]		&	7.71(1)			&	9.47(1)			&	10.03(1)		&	7.30(1)			&	9.28(1)			&	9.59(1)			\\
	AVQZ-BFD	&	exFCI\fnm[3]		&	7.63[-0.05]		&	9.43[-0.03]		&	9.97[-0.06]		&	7.21[-0.09]		&	9.26[-0.03]		&	9.52[-0.06]		\\
				&	SHCI\fnm[2]			&					&					&	9.97(2)[-0.05]	&					&					&					\\
				&	exDMC\fnm[3]		&	7.65(1)[-0.06]	&	9.45(1)[-0.02]	&	10.02(1)[-0.01]	&	7.22(1)[-0.08]	&	9.24(1)[-0.04]	&	9.52(1)[-0.07]	\\
				&	DMC\{J,O\}\fnm[2]	&					&					&	10.01(1)		&					&					&					\\
\hline
	CBS			&	exFCI\fnm[1]		&	7.70			&	9.48			&	10.03			&	7.31			&	9.30			&	9.58			\\
				&	exDMC\fnm[1]		&	7.70(1)			&	9.46(1)			&	10.01(1)		&	7.30(1)			&	9.28(1)			&	9.57(1)			\\
	CBS-BFD		&	exFCI\fnm[3]		&	7.65[-0.05]		&	9.46[-0.02]		&	9.98[-0.05]		&	7.24[-0.07]		&	9.28[-0.02]		&	9.52[-0.06]		\\
				&	exDMC\fnm[3]		&	7.66(1)[-0.04]	&	9.48(1)[+0.02]	&	10.04(1)[+0.03]	&	7.23(1)[-0.07]	&	9.27(1)[-0.01]	&	9.53(1)[-0.04]	\\
\hline
	TBE\fnm[4]	&						&	7.70			&	9.47			&	9.97			&	7.33			&	9.30			&	9.59		\\
	Exp.\fnm[5]	&						&	7.41			&	9.20			&	9.67			&	7.20			&	8.90			&	9.46		\\
	
	\end{tabular}
\end{ruledtabular}
\fnt[1]{Reference \onlinecite{Scemama_2018b}.}
\fnt[2]{Reference \onlinecite{Blunt_2018}.}
\fnt[3]{This work.}
\fnt[4]{Theoretical best estimates of Ref.~\onlinecite{Loos_2018b} obtained from exFCI/AVQZ data corrected with the difference between CC3/AVQZ and CC3/d-aug-cc-pV5Z values.}
\fnt[5]{Energy loss experiment from Ref.~\citenum{Ralphs_2013}.}
\end{table*}
\end{squeezetable}

\section{CIPSI for excited states}
\label{sec:CIPSI}
As mentioned above, our sCI method is based on the CIPSI algorithm. \cite{Huron_1973}
For a calculation involving $\Nst$ states, the CIPSI algorithm, represented in Fig.~\ref{fig:algo}, starts with the following wave functions 
\begin{equation}
	\ket*{\Psi_k^{(0)}}= \sum_{I \in \mD_0} c_{I,k}^{(0)} \kI,
\end{equation}
where $0 \le k \le \Nst-1$.
For a ground-state calculation, $\mD_0$ is usually taken as the HF determinant only, or a determinant made of natural orbitals obtained from a preliminary calculation.
The second option usually significantly speeds up the convergence to the FCI limit.
In the case of an excited-state calculation, $\mD_0$ contains the HF determinant as well as all single excitations (CIS wave function) and state-averaged natural orbitals are usually employed.

Then, we enter the CIPSI iterative process and look for the set $\mA_i$ of (external) determinants $\ka$ connected to the set $\mD_i$ of (internal) determinants $\kI$, i.e.~$\mel*{\alpha}{\hH}{I} \neq 0$.

Next, following Angeli and Persico, \cite{Angeli_1997} we calculate, using Epstein-Nesbet perturbation theory, the second-order energy contribution for each determinant $\ka$ averaged over all states 
\begin{equation}
\label{eq:dE}
	 \delta E(\alpha) = \sum_{k}^{\Nst} \frac{c_{\alpha k}}{\max_I{c_{I k}^2}} \mel*{\Psi^{(i)}_{k}}{\hH}{\alpha},
\end{equation}
with
\begin{equation}
	c_{\alpha k} = \frac{\mel*{\Psi^{(i)}_k}{\hH}{\alpha}}{\mel*{\Psi^{(i)}_k}{\hH}{\Psi^{(i)}_k} - \mel*{\alpha}{\hH}{\alpha}}.
\end{equation}
This choice gives a balanced selection between states of different multi-configurational nature.
We then select the determinants $\ket*{\alpha^*}$ having the largest contributions, i.e.
\begin{equation}
	 \delta E(\alpha^*) = \max_{\alpha \in \mA_i} \delta E(\alpha).
\end{equation}
The subset $\mA_i^* \subset \mA_i$ of determinants $\ket*{\alpha^*}$ are then added to $\mD_i$ to form $\mD_{i+1}$, i.e.~$\mD_{i+1} = \mD_i \cup \mA_i^*$.

This process is repeated until convergence of the ground- and excited-state energies given by the lowest eigenvalues of the Hamiltonian $\hH$.
At convergence, the CIPSI algorithm provides ground- and excited-state wave functions
\begin{equation}
\label{eq:Psi4QMC}
	\ket*{\Psi_k^{(n)}}= \sum_{I \in \mD_n} c_{I,k} \kI
\end{equation}
that can be used for QMC calculations.

\section{Computational details}
\label{sec:comp_details}
The sCI calculations have been performed with the electronic structure software \textsc{quantum package}, \cite{Garniron_2019} while the QMC calculations have been performed with the \textsc{qmc=chem} program. \cite{qmcchem, Scemama_2013} 
Both software packages are developed in Toulouse and are freely available.
Our computational procedure follows closely the one reported in Ref.~\onlinecite{Scemama_2018b}, where the interested reader will find additional details about trial wave functions and our Jastrow-free QMC protocol.
Below, we report more information regarding pseudopotentials.
The ground state geometry of \ce{H2O} has been obtained at the CC3/aug-cc-pVTZ level without frozen core approximation. 
This geometry has been extracted from Ref.~\onlinecite{Loos_2018b} and is also reported as {\SI} for sake of completeness.
The sCI calculations have been performed in the frozen-core approximation with the CIPSI algorithm \cite{Huron_1973} which selects perturbatively determinants in the FCI space. \cite{Giner_2013, Giner_2015, Caffarel_2014, Scemama_2014, Caffarel_2016, Scemama_2016, Garniron_2017b, Scemama_2018a, Scemama_2018b, Loos_2018b, Dash_2018, Loos_2019}

For the calculations involving pseudopotentials, we have used the valence-only Burkatzki-Filippi-Dolg (BFD) cc-pVXZ basis sets (with X $=$ D, T and Q) in conjunction with the corresponding BFD small-core pseudopotentials. \cite{Burkatzki_2007, Burkatzki_2008} 
The diffuse functions from the standard (all-electron) Dunning basis set family aug-cc-pVXZ were then added to the (diffuseless) BFD bases.
In the following, we labeled as AVXZ and AVXZ-BFD the all-electron Dunning and valence-only BFD bases, respectively.

The FN-DMC simulations are performed using the stochastic reconfiguration algorithm developed by Assaraf et al., \cite{Assaraf_2000} with a time-step of $2 \times 10^{-4}$ au.
In the present case, it is not necessary to perform time step extrapolations as the time step error is smaller than the statistical error in the computation of excitation energies.
Preliminary calculations have shown that using the T-moves scheme in FN-DMC \cite{Casula_2006,Casula_2010} had no influence in the calculation of the excitation energies.
This observation is in agreement with the recent results of Blunt and Neuscamman on the same system. \cite{Blunt_2019}
As pointed out by Hammond and coworkers, \cite{Hammond_1987} when the trial wave function does not include a Jastrow factor, the non-local pseudopotential can be localized analytically and the usual numerical quadrature over the angular part of the non-local pseudopotential can be eschewed. 
In practice, the calculation of the localized part of the pseudopotential represents only a small overhead (about 15\%) with respect to a calculation without pseudopotentials (and the same number of electrons).
For more details about our implementation of pseudopotentials within QMC, we refer the interested readers to Ref.~\citenum{Caffarel_2016b}.

\begin{figure*}
	\includegraphics[width=0.8\linewidth]{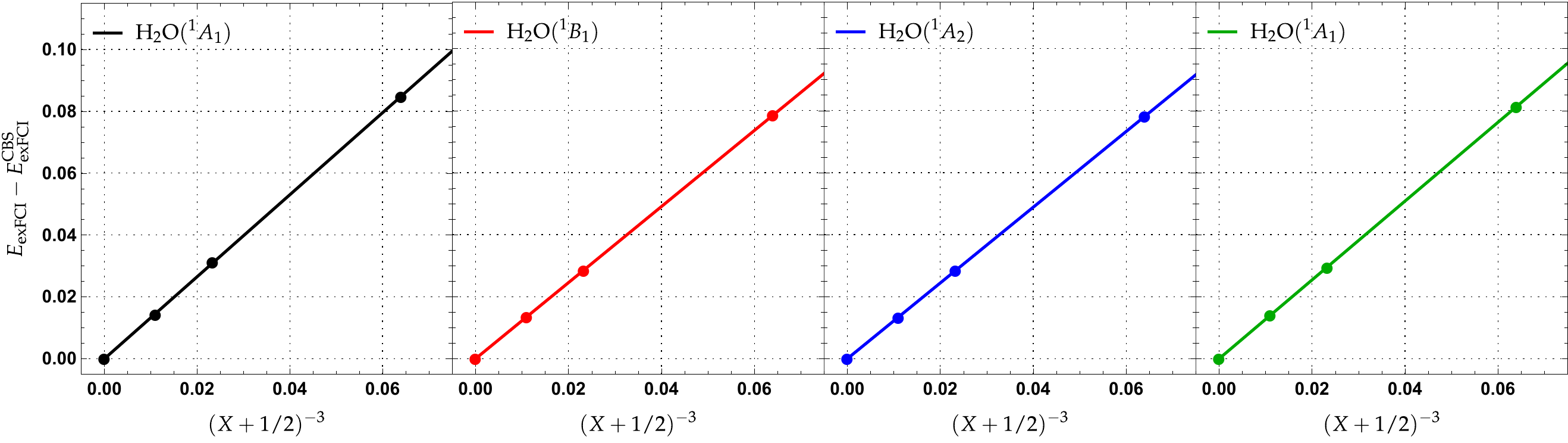}
	\includegraphics[width=0.6\linewidth]{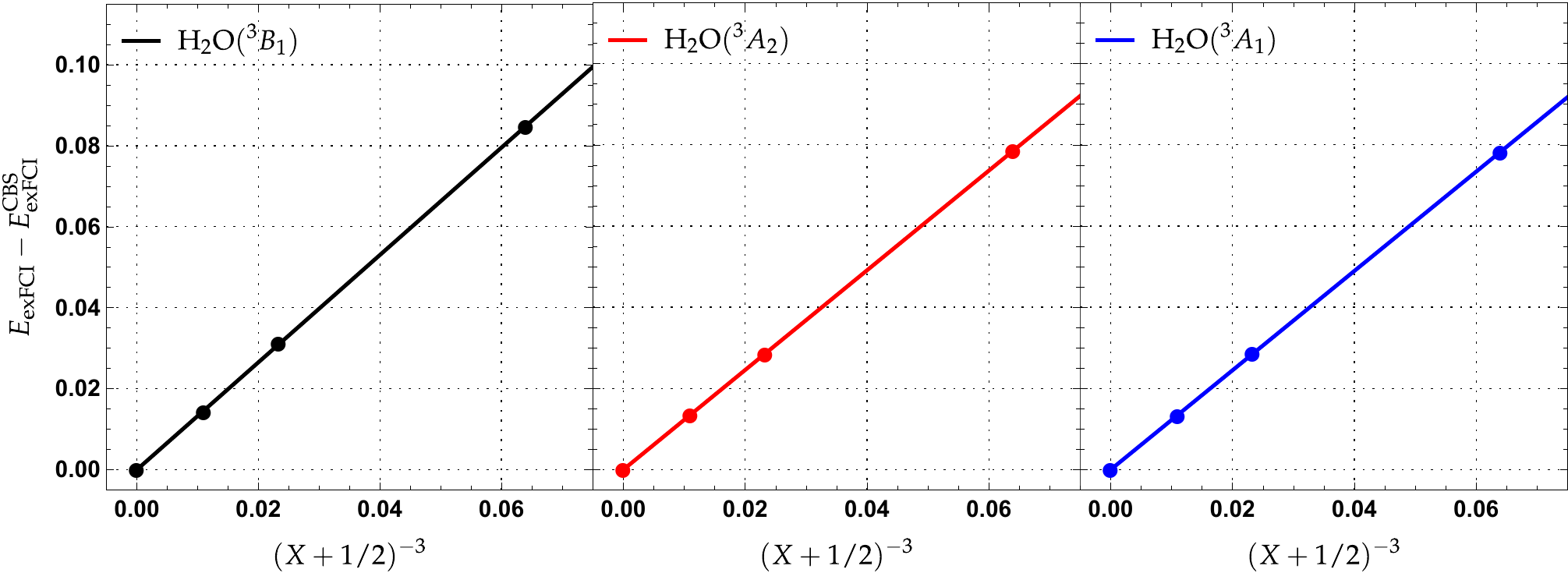}
	\caption{
	\label{fig:H2O_CI_CBS}
	Extrapolation of the exFCI energies to the complete basis set (CBS) limit for the water molecule.
	The extrapolated sCI energy $\EexFCI$ is plotted as a function of $(\tX+1/2)^{-3}$ for $\tX = 2$ (AVDZ-BFD), $\tX = 3$ (AVTZ-BFD) and $\tX = 4$ (AVQZ-BFD).
	$\EexFCI^\text{CBS}$ stands for the CBS energy obtained at the exFCI level.
	}
\end{figure*}

\section{Results}
\label{sec:res}

\begin{squeezetable}
\begin{table}
\caption{
\label{tab:H2O-BFD-DMC}
Vertical excitation energies (in eV) for the three lowest singlet and three lowest triplet excited states of water obtained with the BFD pseudopotentials and the valence-only AVXZ basis sets (X $=$ D, T, and Q). 
$\Ndet$ is the number of determinants in the trial wave functions.} 
\begin{ruledtabular}
\begin{tabular}{lrdrdrd}
Transition					& \multicolumn{2}{c}{AVDZ-BFD} 	& \multicolumn{2}{c}{AVTZ-BFD}		& \multicolumn{2}{c}{AVQZ-BFD}	\\
								\cline{2-3}							\cline{4-5}							\cline{6-7}					
						
	 						&	$\Ndet$ 		&	\tabc{FN-DMC} 		&	$\Ndet$ 		&	\tabc{FN-DMC} 		&	$\Ndet$		&	\tabc{FN-DMC}  			\\
\hline
\exc{1}{B}{1}{}				&	$8\,825$		&	7.67(1)	&	$8\,655$		&	7.68(1)	&	$8\,856$		&	7.71(1)	\\
 							&	$65\,600$		&	7.66(1)	&	$82\,387$		&	7.67(1)	&	$97\,937$		&	7.68(1)	\\
							&	$287\,688$		&	7.65(1)	&	$334\,839$		&	7.66(2)	&	$532\,734$		&	7.69(1)	\\
							&	$646\,643$		&	7.65(1)	&	$694\,560$		&	7.67(1)	&	$1\,579\,987$	&	7.63(1)	\\
							&	\bf exDMC			&	7.65(1)	&					&	7.66(1)	&					&	7.65(1)	\\
						\hline
\exc{1}{A}{2}{}				&	$8\,825$		&	9.46(1)	&	$8\,655$		&	9.49(1)	&	$8\,856$		&	9.47(1)	\\
							&	$65\,600$		&	9.45(1)	&	$82\,387$		&	9.47(1)	&	$97\,937$		&	9.48(1)	\\
							&	$287\,688$		&	9.45(1)	&	$334\,839$		&	9.50(2)	&	$532\,734$		&	9.49(1)	\\
							&	$646\,643$		&	9.45(1)	&	$694\,560$		&	9.47(1)	&	$1\,579\,987$	&	9.44(1)	\\
							&	\bf exDMC			&	9.45(1)	&					&	9.49(1)	&					&	9.45(1)	\\
						\hline
\exc{1}{A}{1}{}				&	$8\,825$		&	10.05(1)	&	$8\,655$		&	10.07(1)	&	$8\,856$		&	10.08(1)	\\
							&	$65\,600$		&	10.03(1)	&	$82\,387$		&	10.03(1)	&	$97\,937$		&	10.04(1)	\\
							&	$287\,688$		&	10.01(1)	&	$334\,839$		&	10.02(2)	&	$532\,734$		&	10.04(1)	\\
							&	$646\,643$		&	10.00(1)	&	$694\,560$		&	10.04(1)	&	$1\,579\,987$	&	10.01(1)	\\
							&	\bf exDMC			&	10.00(1)	&					&	10.04(1)	&					&	10.02(1)	\\
						\hline
\exc{3}{B}{1}{}				&	$5\,848$		&	7.23(1)	&	$6\,532$		&	7.25(1)	&	$6\,446$		&	7.25(1)	\\
							&	$51\,538$		&	7.24(1)	&	$68\,255$		&	7.24(1)	&	$70\,637$		&	7.23(1)	\\
							&	$289\,748$		&	7.25(1)	&	$473\,245$		&	7.23(1)	&	$424\,318$		&	7.24(1)	\\
							&	$1\,518\,066$	&	7.28(1)	&	$2\,128\,116$	&	7.25(1)	&	$1\,695\,420$	&	7.21(1)	\\
							&	\bf exDMC			&	7.26(1)	&					&	7.25(1)	&					&	7.22(1)	\\
						\hline
\exc{3}{A}{2}{}				&	$5\,848$		&	9.23(1)	&	$6\,532$		&	9.26(1)	&	$6\,446$		&	9.25(1)	\\
							&	$51\,538$		&	9.29(1)	&	$68\,255$		&	9.28(1)	&	$70\,637$		&	9.28(1)	\\
							&	$289\,748$		&	9.29(1)	&	$473\,245$		&	9.29(1)	&	$424\,318$		&	9.28(1)	\\
							&	$1\,518\,066$	&	9.25(1)	&	$2\,128\,116$	&	9.29(2)	&	$1\,695\,420$	&	9.23(1)	\\
							&	\bf exDMC			&	9.27(1)	&					&	9.30(1)	&					&	9.24(1)	\\
						\hline
\exc{3}{A}{1}{}				&	$5\,848$		&	9.54(1)	&	$6\,532$		&	9.54(1)	&	$6\,446$		&	9.54(1)	\\
							&	$51\,538$		&	9.55(1)	&	$68\,255$		&	9.53(1)	&	$70\,637$		&	9.54(1)	\\
							&	$289\,748$		&	9.54(1)	&	$473\,245$		&	9.54(1)	&	$424\,318$		&	9.54(1)	\\
							&	$1\,518\,066$	&	9.54(1)	&	$2\,128\,116$	&	9.53(1)	&	$1\,695\,420$	&	9.50(1)	\\
							&	\bf exDMC			&	9.54(1)	&					&	9.55(1)	&					&	9.52(1)	\\
\end{tabular}
\end{ruledtabular}
\end{table}
\end{squeezetable}

\subsection{Selected configuration interaction}
\label{sec:sCI}
Vertical excitation energies for various singlet and triplet states of the water molecule are reported in Table \ref{tab:H2O}.
For a molecule as small as water (even in a fairly large basis set), it is straightforward to converge sCI calculations and to obtain vertical excitation energies with an uncertainty (for a given basis) of \IneV{0.01}.
Throughout the paper, we label these calculations as exFCI (extrapolated FCI) for consistency with our previous studies. \cite{Scemama_2018a, Scemama_2018b, Loos_2018b, Loos_2019}
In Table \ref{tab:H2O}, the relative difference between the all-electron and the corresponding BFD pseudopotential calculations is reported in square brackets.
For comparison, we also report the (extrapolated) energies of Blunt and Neuscamman \cite{Blunt_2019} obtained with the semistochastic heat-bath CI (SHCI) method, \cite{Holmes_2016, Sharma_2017,Li_2018} one of the other sCI variants.
As expected, these values agree perfectly (within statistical error) with the exFCI energies.

Table \ref{tab:H2O} also contains complete basis set (CBS) estimates obtained with the usual extrapolation formula \cite{HelgakerBook}
\begin{equation}
	\EexFCI(\tX) = \EexFCI^\text{CBS} + \frac{\alpha}{(\tX+1/2)^{3}},
\end{equation}
where $\alpha$ and $\EexFCI^\text{CBS}$ are obtained by fitting the exFCI results for $\tX = 2$ (AVDZ), $\tX = 3$ (AVTZ), and $\tX = 4$ (AVQZ).
For the BFD bases, these fits are represented in Fig.~\ref{fig:H2O_CI_CBS} for the four singlet and three triplet transitions studied here.
The corresponding all-electron extrapolations can be found in Ref.~\onlinecite{Scemama_2018b}.
From Fig.~\ref{fig:H2O_CI_CBS}, it is clear that these extrapolations can be safely trusted.

At the sCI level, one can clearly see that, for both spin manifolds, the BFD pseudopotentials induce a rather systematic redshift on the excitation energies of magnitude \IneV{0.05} (i.e.~roughly 1 kcal/mol) which may or may not be an acceptable error depending on the target accuracy.
The maximum error is found to be \IneV{-0.09} for the first triplet state whereas the minimum errors are as small as \IneV{0.02--0.03} in some cases.

\begin{figure*}
	\includegraphics[width=0.8\linewidth]{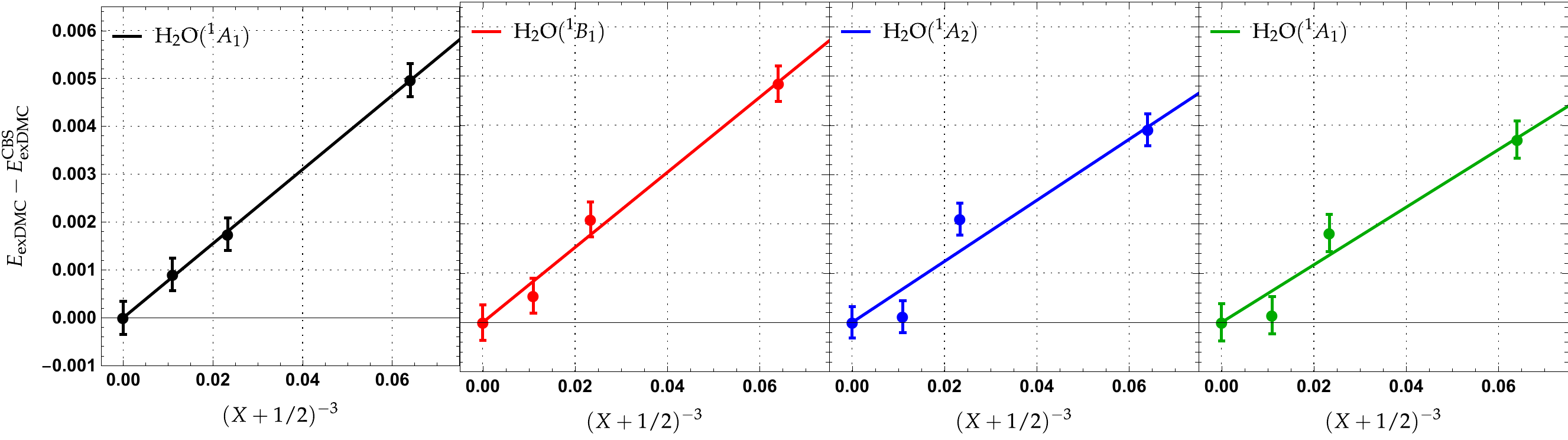}
	\includegraphics[width=0.6\linewidth]{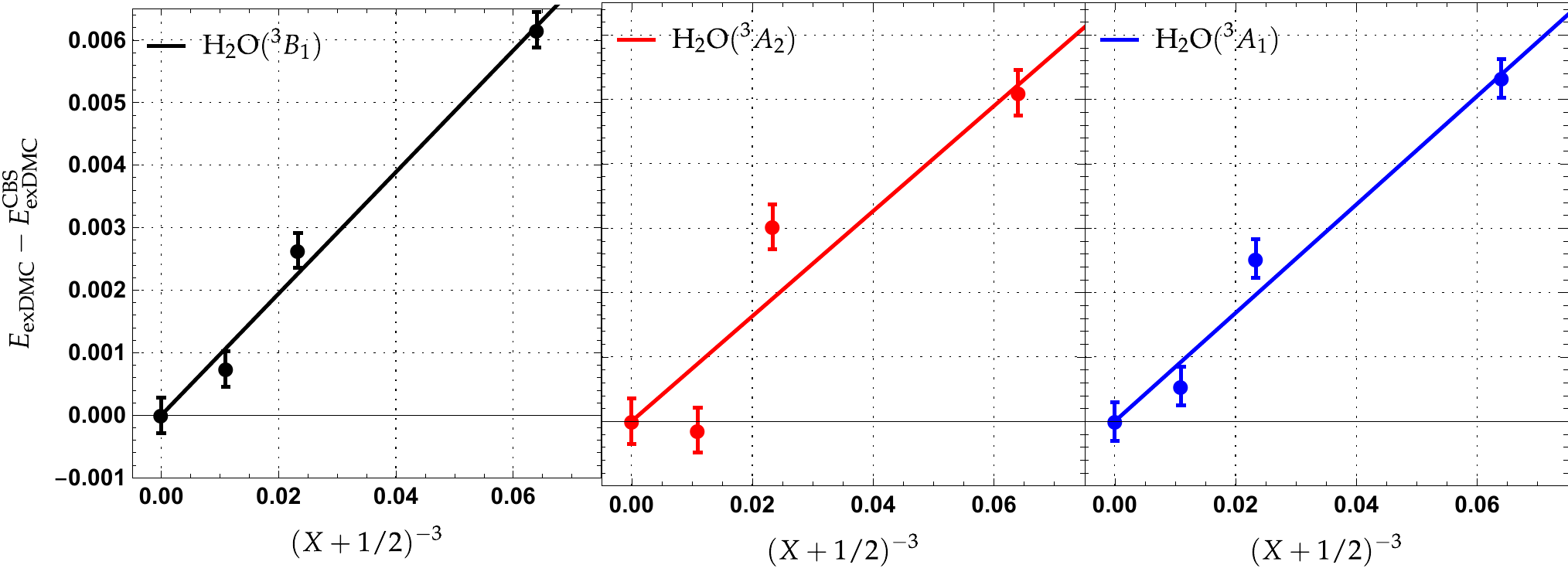}
	\caption{
	\label{fig:H2O_CBS}
	Extrapolation of the exDMC energies to the complete basis set (CBS) limit for the water molecule.
	The extrapolated FN-DMC energy $\EexDMC$ is plotted as a function of $(\tX+1/2)^{-3}$ for $\tX = 2$ (AVDZ-BFD), $\tX = 3$ (AVTZ-BFD) and $\tX = 4$ (AVQZ-BFD).
	$\EexDMC^\text{CBS}$ stands for the CBS energy obtained at the exDMC level.
	}
\end{figure*}

\subsection{Diffusion Monte Carlo}
\label{sec:DMC}

Our ultimate goal is to obtain the FN-DMC energies associated with the FCI wave functions.
However, the ground- and excited-state FCI wave functions are obviously too large to be used as trial wave functions in FN-DMC calculations.
Therefore, we use truncated CIPSI expansions (generated as explained in Sec.~\ref{sec:CIPSI}) of increasing lengths as trial wave functions, and extrapolations are performed in order to estimate the FN-DMC energies one would obtain with the FCI wave functions.
In Table \ref{tab:H2O-BFD-DMC}, we report the singlet and triplet excitation energies of water obtained at the FN-DMC level for various multideterminantal trial wave functions 
\begin{equation}
	\label{eq:PsiT}
	\PsiT = \sum_{I}^{\Ndet} c_{I} \kI
\end{equation}
of size $\Ndet$ and variational energy $\EsCI$ (where $\kI$ is a Slater determinant and $c_{I}$ its corresponding CI coefficient).
The extrapolated FN-DMC results, labeled as exDMC and reported in Table \ref{tab:H2O}, are obtained by performing a linear extrapolation of the FN-DMC energy $\EDMC$ as a function of $\EexFCI-\EsCI$ for various values of $\Ndet$.
Identifying the quantity $\EexFCI-\EsCI$ as the variational bias introduced by the truncation of the trial wave function, based on these smaller trial wave functions, we can extrapolate $\EDMC$ to $\EexFCI-\EsCI = 0$ in order to estimate the FN-DMC energy of the FCI trial wave function.
Additional details about this procedure can be found in Refs.~\onlinecite{Scemama_2018a, Scemama_2018b, Loos_2018b}.
The graphs associated with these extrapolations are reported as {\SI} for the singlet and triplet transitions.
It is noteworthy that only the last three points are taken into account in the linear extrapolation, i.e., the point corresponding to the smallest trial wave function is systematically discarded.

Following a similar procedure as for exFCI (see Sec.~\ref{sec:sCI}), we have performed CBS extrapolations of the exDMC energies.
These are represented in Fig.~\ref{fig:H2O_CBS}.
At first sight, it seems that the CBS extrapolations of the exDMC energies are less trustworthy than their variational versions (see Fig.~\ref{fig:H2O_CI_CBS}).
However, it is important to realize that there is a factor of about 16 between the energy scale of the two extrapolation sets in Figs.~\ref{fig:H2O_CI_CBS} and \ref{fig:H2O_CBS}.
In other words, the exDMC extrapolation lines are much flatter than their exFCI counterparts, which does explain their magnified sensitivity.
For extra statistics, the two sets of energies can be used altogether as they must extrapolate to the same CBS limit.

At this state, it is worth emphasizing that it is particularly reassuring that, in most cases, the excitation energies obtained at the exFCI and exDMC levels do converge, within statistical error, to the same CBS limit (that is, the exact energy) as it should be. 
This key observation validates the here-proposed strategy for the CBS extrapolation.
However, there is one case for which it is not true, namely the \ex{1}{A}{1}{}{n}{3s} transition, where $\EexFCI^\text{CBS}$ and $\EexDMC^\text{CBS}$ are significantly different (\IneV{0.06}).
This can be explained by the particularly strong basis set effect associated with the pronounced Rydberg nature of this transition.
Indeed, we have recently shown that, even within conventional deterministic wave function methods such as high-level coupled cluster theories, this particular state requires doubly-augmented basis sets (d-aug-cc-pVXZ) to be properly modeled. \cite{Loos_2018b}

Compared to the conclusion drawn in Sec.~\ref{sec:sCI}, the excitation energies gathered in Table \ref{tab:H2O} show that the deviation between the all-electron and valence-only results are slightly larger at the FN-DMC level.
Yet, this discrepancy is fairly acceptable for usual chemical applications with a maximum error of \IneV{0.07}, especially knowing the inherent uncertainties associated with stochastic simulations.
In this regard, we can point out that the excitation energies of Blunt and Neuscamman (obtained with their simple two-determinant ansatz labeled as DMC\{J,O\} in Table \ref{tab:H2O}) seem to benefit from small, yet systematic, error compensations. \cite{Blunt_2019}

As a final remark, we would like to point out that, in a large number of cases, we see that the difference between all-electron and pseudopotential calculations can be transferred from the variational to the FN-DMC level.
Consequently, if one is able to estimate the error induced by the pseudopotentials at the sCI level, it should provide a reasonable estimate of the error that should occur in the FN-DMC excitation energies.

\section{Conclusion}
\label{sec:ccl}
In the present manuscript, we have reported a preliminary study on the influence of BFD pseudopotentials (and their corresponding basis sets) on vertical excitation energies obtained at the FN-DMC level with a Jastrow-free protocol.
By comparing valence-only and all-electron calculations performed for six low-lying states of the water molecule, we clearly evidence that a small and systematic error is induced by the pseudopotentials and their associated basis set: the transition energy is red-shifted by \IneV{0.05} at the variational level and slightly more at the FN-DMC level.
The similarity between the variational and FN-DMC shifts hints that most of the localization error associated with the use of pseudopotentials cancels out to a large extent when one computes excitation energies.
Hence, the discrepancies between all-electron and valence-only calculations might originate mainly from the difference in the one-electron basis sets.
Overall, the small bias introduced by the BFD pseudopotentials and basis sets is acceptable for the vast majority of applications, but could be problematic when looking for very high precision (like in benchmark studies).
Finally, we would like to mention that it would be particularly interesting and instructive to test the new generation of pseudopotentials developed by Mitas and coworkers. \cite{Bennet_2017}

\section*{Supplementary material}
\label{sec:SI}
See {\SI} for the geometry of the water molecule and the graphs associated with the DMC extrapolations.

\begin{acknowledgments}
PFL would like to thank Eric Neuscamman for valuable discussions.
Funding from \emph{Projet International de Coop\'eration Scientifique} (PICS08310) is acknowledged.
This work was performed using HPC resources from CALMIP (Toulouse) under allocation 2019-18005 and from GENCI-TGCC (Grant 2018-A0040801738).
AB was supported by the U.S.~Department of Energy, Office of Science, Basic Energy Sciences, Materials Sciences and Engineering Division, as part of the Computational Materials Sciences Program and Center for Predictive Simulation of Functional Materials.
\end{acknowledgments}

\bibliography{ECP}

\end{document}